\documentclass[aps,twocolumn,showpacs]{revtex4-1}
\usepackage{dcolumn}
\usepackage{bm}
\usepackage{amssymb}
\usepackage{amsfonts}
\usepackage{amsmath}
\usepackage{graphicx}
\begin{document}

\title{The long-tail distribution function of mutations in bacteria}
\author{Augusto Gonzalez}
\affiliation{Instituto de Cibernetica, Matematica y Fisica, La Habana, Cuba}

\begin{abstract}
Levy flights in the space of mutations model time evolution of bacterial DNA. Parameters in the
model are adjusted in order to fit observations coming from the Long Time Evolution Experiment
with E. Coli.
\end{abstract}

\maketitle

{\bf The Long Time Evolution Experiment.}
I recall the extremely interesting experiment with E. Coli, conducted by Prof. R. Lenski and his 
group \cite{ref1,ref2}, and running already for more than 27 years. Among the reported results, 
I use the following \cite{ref3}:

1. In a culture of bacteria, after 20,000 generations, around $3\times 10^8$ single point mutations
in the DNA are registered. These are local modifications of the DNA chain. I notice that the 
number of bacteria undergoing continuous evolution is around $5\times 10^6$.

2. They measure also the frequency of mutations involving rearrangements in segments of the DNA.
In particular, mutations in which the repair mechanisms are damaged and the mutation rate increases
100 times. This mutator phenotype becomes dominant in two of twelve cultures (probability 1/6)
after 2500 - 3000 generations, in a third culture (cumulative probability 1/4) after 8,500 
generations, and in a fourth culture (cumulative probability 1/3) after 15,000 generations.

The purpose of my paper is to present a model for mutations in bacteria and to adjust the
model parameters in order to qualitatively fit these data.

{\bf The accumulative character of mutations.}

\begin{figure}[ht]
\begin{center}
\includegraphics[width=0.9\linewidth,angle=0]{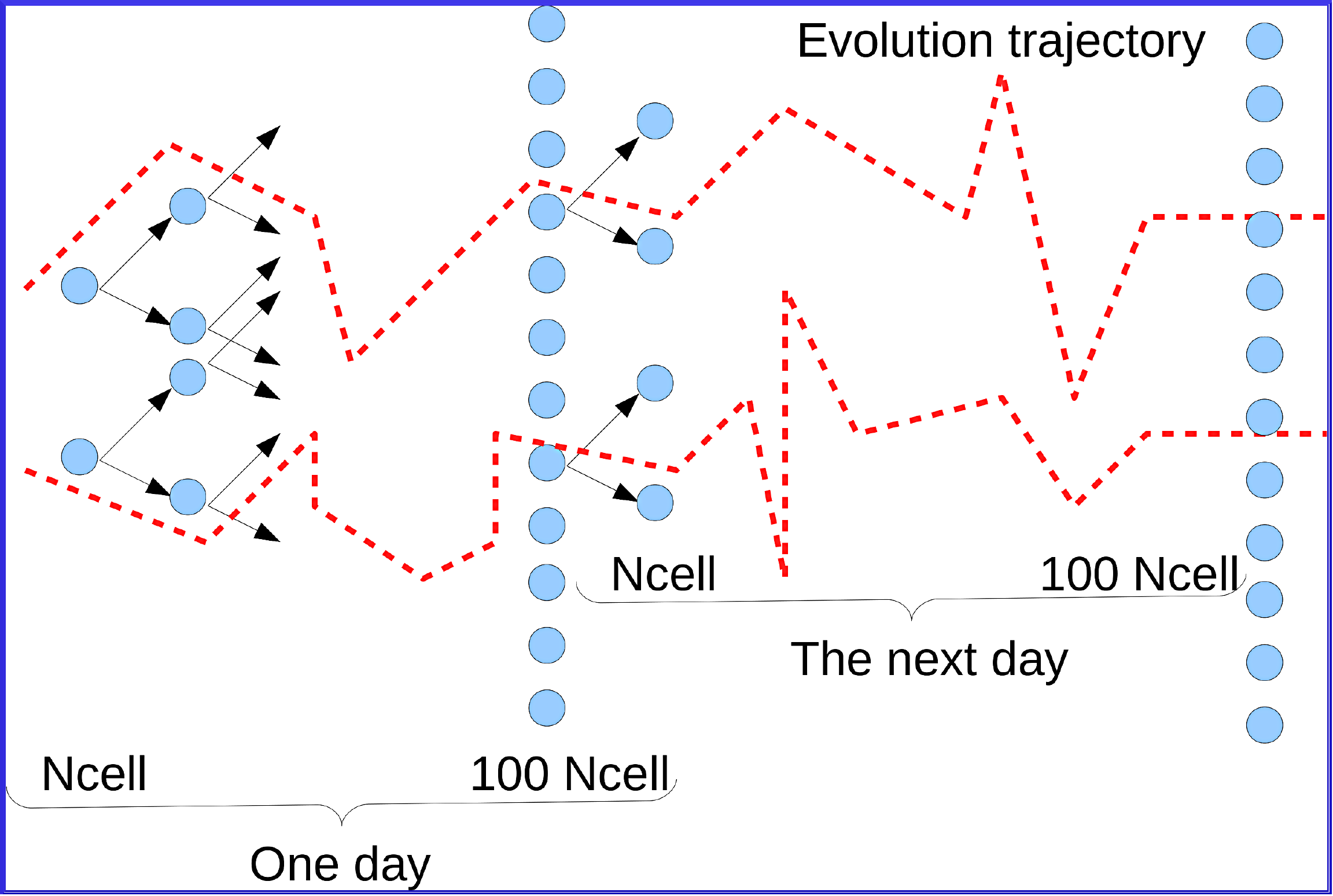}
\caption{Schematic representation of the evolution of bacteria in the Long Time Evolution Experiment. 
Every day, the cells experience a clonal expansion in which the initial number $N_{cell}\approx 5\times
10^6$ is raised 100 times. However, only $N_{cell}$ bacteria pass to the next day. The evolution 
trajectories of two cells are marked by red dashed lines.}
\label{fig1}
\end{center}
\end{figure}

In my model, the time evolution of cells defines trajectories, 
as schematically represented in Fig. \ref{fig1}, where two of these trajectories are
drawn in red.

The idea about trajectories in the evolution of cells means that there are Markov chains \cite{ref4}
of mutations, where the change in the DNA of a cell at step $i+1$, $x_{i+1}$, comes from the 
change in the previous step plus an additional modification:

\begin{equation}
x_{i+1}=x_{i}+\delta
\label{eq1}
\end{equation}

\noindent
Horizontal DNA transfer is not considered.

{\bf Measuring changes in the DNA}
A single strand of E. Coli DNA contains around $4.6\times10^6$ bases of a four letter alphabet:
A, G, C, and T. \cite{ref5} In order to measure changes in the DNA, one may use a variable similar
to that one of paper \cite{ref6}.

First, define an auxiliary variable at site $\alpha$ in the molecule: $u_{\alpha}(G)=3/8$,
$u_{\alpha}(A)=1/8$, $u_{\alpha}(T)=-1/8$, and $u_{\alpha}(C)=-3/8$. Then, define a walk 
along the DNA:

\begin{equation}
y(\beta)=\sum_{\alpha=1}^{\beta}u_{\alpha}.
\label{eq2}
\end{equation}

As a function of $\beta$, the variable $y$ draws a profile of the DNA molecule, and 
modifications can be measured as: $X(\beta)=y(\beta)-y_0(\beta)$. 
where $y$ correspond to the mutated DNA, and $y_0$ -- to the initial configuration. 
Of course, there are so many $X(\beta)$,
five millions, that they are not of practical use. The strategy could be to use variables 
measuring global changes or distances to the original function:

\begin{equation}
X=\sum_{\alpha=1}^L (u'_{\alpha}-u_{\alpha}),
\label{eq3}
\end{equation}

\begin{equation}
X^{(1)}=\sum_{\alpha=1}^L \alpha(u'_{\alpha}-u_{\alpha}),
\label{eq4}
\end{equation}

\noindent
$X^{(2)}$ (the second moment), etc. $L$ is the length of the molecule. The Shannon
informational entropy \cite{ref7} could also be of use.

In what follows, I shall assume that mutations are well characterized by a few global
variables.

{\bf Levy model of mutations}
The $\delta$ term in Eq. (\ref{eq1}) represents mutations at step $i+1$. It may come from a 
partially repaired damage in the DNA that is fixed after replication, or from a prune 
error in the replication process. It should be stressed that both the repair mechanisms
and the replication process guarantee very high fidelities. The error introduced by the 
latter, for example, is around one mistaken base per $10^9$ bases in the human DNA strand \cite{ref8}.

\begin{figure}[ht]
\begin{center}
\includegraphics[width=0.9\linewidth,angle=0]{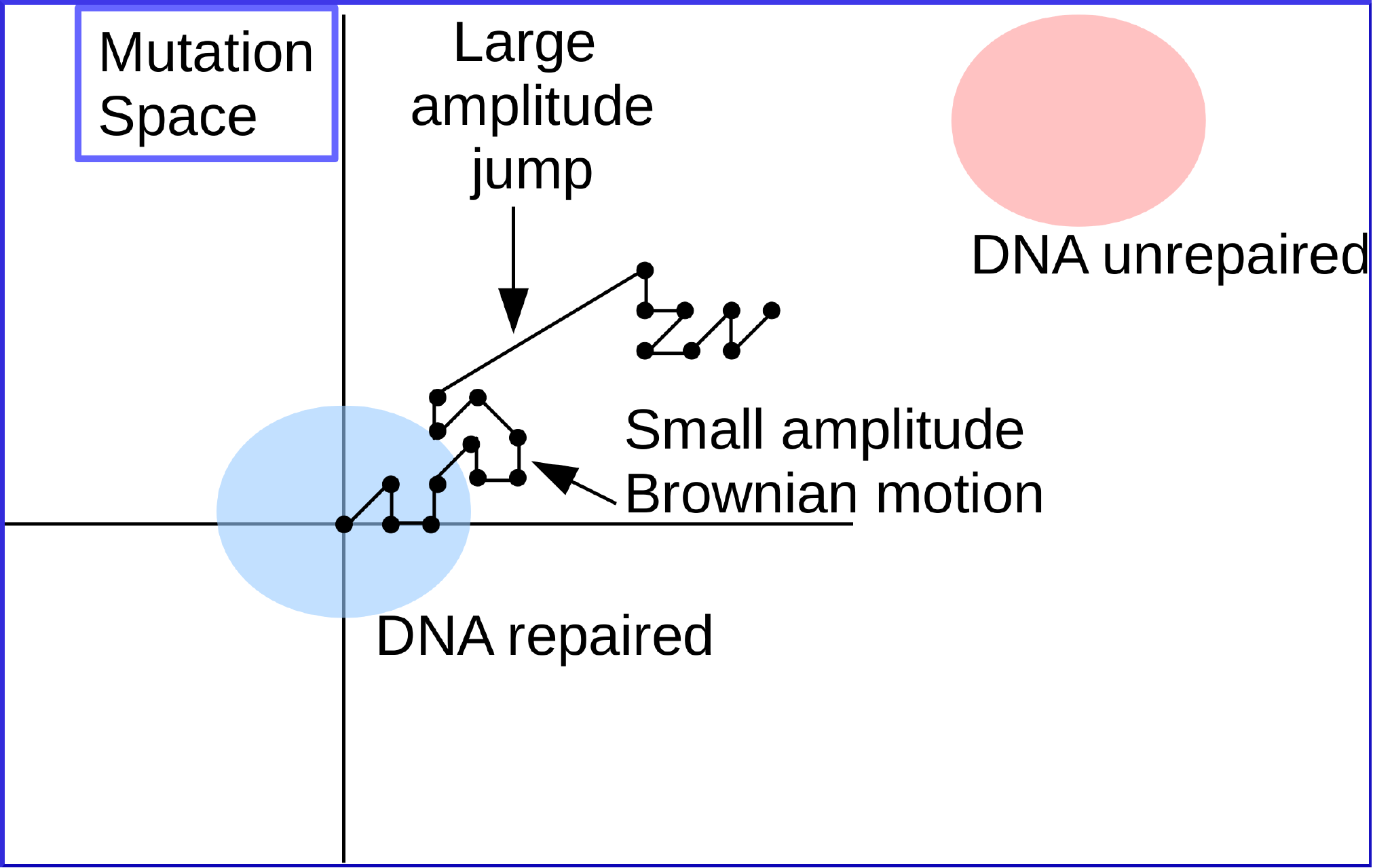}
\caption{Schematic representation of a single cell mutation trajectory. The starting point
is $X=0$. In the mutation space, I distinguished regions in which the DNA repair mechanism
is active or damaged.}
\label{fig2}
\end{center}
\end{figure}

Let me stress once again that $\delta$ is not the damage caused by endogenous or external
factors, but the resulting modification after the action of the repair mechanisms. It is known, 
for example, that ionizing radiation may cause double strand breaks in the DNA \cite{ref9}. These
damages are very difficult to repair \cite{ref8}. The repair mechanism itself may introduce large
changes in the resulting DNA composition after a double strand break event.

My proposal for $\delta$ is the following: $\delta=\delta_B+\delta_{LJ}$. The $\delta_B$
component corresponds to a Brownian motion with maximal amplitude $D_B$. 
Notice that $D_B=1$ would mean roughly a change of basis in each replication step because
$u_{\alpha}(G)-u_{\alpha}(C)=3/4$. This Brownian motion introduces
local modifications in the DNA. After $N_{step}$ replication steps, the characteristic 
dispersion of a trajectory due to this Brownian motion (something like the radius of the 
colored region near the origin in Fig. 2) is $D_B \sqrt{N_{step}}$. \cite{ref10}

The large-jump component of $\delta$, $\delta_{LJ}$, on the other hand, is modeled with
the help of 
rare events with total probability $p << 1$, and a probability density proportional to
$1/\delta_{LJ}^2$, where the amplitude ranges from $D_B$ to infinity (in practice, I will
introduce a cutoff, $D_{max}$). The combination of 
the Brownian motion and the large amplitude jumps leads to Levy flights \cite{ref11} in  
the mutation space, schematically represented in Fig. 2.

Let me notice that the distribution function associated to Levy flights is a fat- or 
long-tail one. This fact could be related to the long range correlations observed in the 
walks along the DNA \cite{ref6}.

{\bf The long-tail distribution function of mutations.}
Four parameters enter my oversimplified Levy model of mutations: $N_{cell}$,
$N_{step}$, $D_B$ and $p$. As mentioned above, $N_{cell}=4.6\times 10^6$. On the other
hand, $N_{step}$ is the number of replication steps along a trajectory.

$D_B$ is the amplitude of the Brownian motion. It shall be determined from the observed
number of single point mutations (SPM) after 20,000 generations. The number of SPMs
per bacteria is $3\times 10^8/(4.6 \times 10^6)\approx 65$. The characteristic 
dispersion of the trajectory, on his side, is the Brownian radius, $\sqrt{N_{step}}~D_B\approx 140~D_B$.
In order to estimate que equivalent number of SPM, I divide the latter by the
mean deviation involved in a SPM, that is 5/12. Notice that $u(G)-u(A)=1/4$, 
$u(G)-u(T)=1/2$, etc. Thus, $65=140~D_B/(5/12)$, and $D_B\approx 0.19$.

Finally, the parameter $p$ is fixed to $1.3\times 10^{-5}$ Below, I shall come back to the way of determining
it. 

In the simulations, all of the  $N_{cell}$ trajectories start at $X=0$. In any
replication step, mutations are given by Eq. (\ref{eq1}),
where $\delta$ contains both the Brownian and the large-amplitude components.

\begin{figure}[ht]
\begin{center}
\includegraphics[width=0.9\linewidth,angle=0]{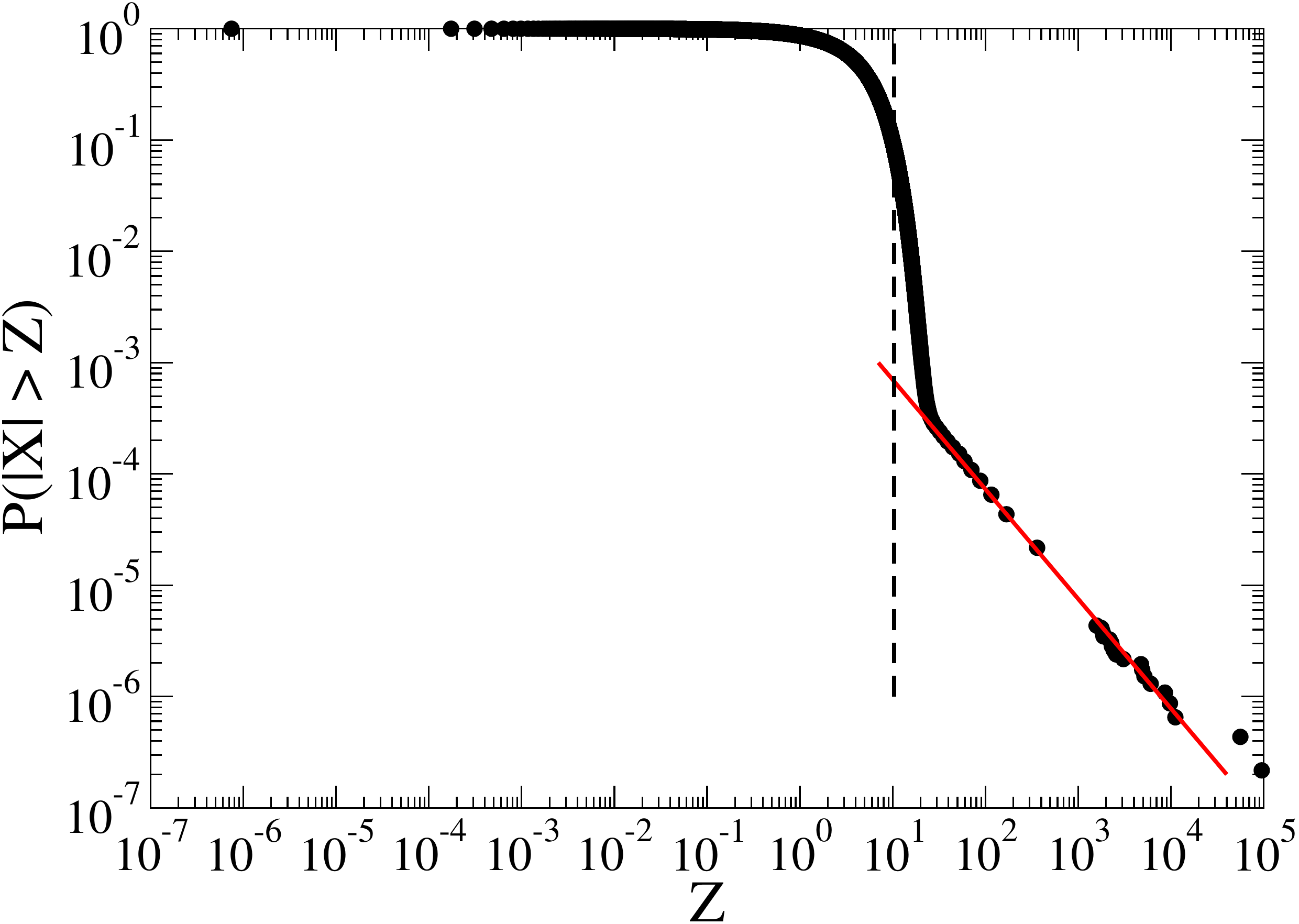}
\caption{The average cumulative probability of mutations, $P(|X|>Z)$, for a single bacterium
after 3000 generations. 
Points come from the numerical simulations, whereas the red solid line is a $1/Z$ fit to 
the tail. The Brownian radius, $D_B \sqrt{N_{step}}$, is marked by a dashed line.}
\label{fig3}
\end{center}
\end{figure}

The probability distribution function for mutations in a cell, $P(X)$, is the probability that
a cell arrives at the end point with an amplitude $X$. For convenience, I compute not $P(X)$, but the
cumulative probability distribution, $P(|X|>Z)$, which is shown in Fig. \ref{fig3} for $N_{step}=3000$.

The Brownian radius, $\sqrt{N_{step}}~D_B\sim 10.4$, concentrating most of
the points, is apparent in the figure. In addition, the tail can be fitted to a
$1/Z$ dependence. The coefficient is roughly $N_{step} D_B~ p$. 

The data on the mutator phenotype is to be used in order to fix the slope in the tail.
I assume that the repair mechanisms are related to a coding region in the DNA 
of length $l$. The mechanisms are damaged when this region suffers modifications
greater than a given $X_{u}$. The cumulative probability can be estimated as 
$N_{cell}~ P(|X|>X_{u})$. Using the functional dependence in the tail, I get:

\begin{equation}
Cum.~Prob. \approx N_{cell}\frac{N_{step}D_B~p}{X_{u}}\frac{l}{L}=a N_{cell}N_{step}.
\label{eq5}
\end{equation} 

\begin{figure}[ht]
\begin{center}
\includegraphics[width=0.9\linewidth,angle=0]{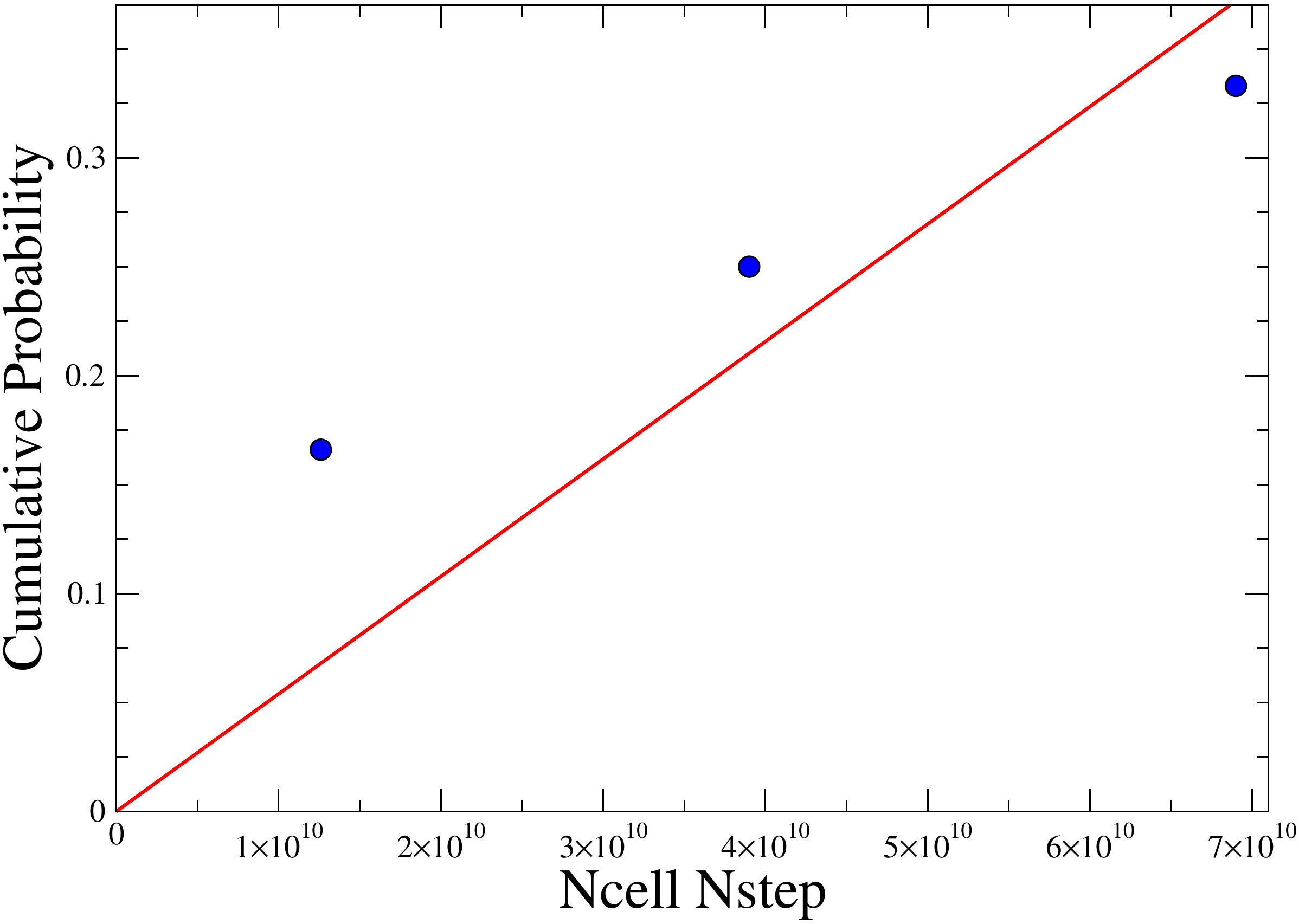}
\caption{Cumulative probability of the mutator phenotype in the Long Time
Evolution Experiment. The line is a fit according to Eq. (\ref{eq5}).}
\label{fig4}
\end{center}
\end{figure}

So far, I do not have precise values for $l$ and $X_{u}$. Reasonable numbers are
$l/L\approx 10^{-2}$, $X_{u}/L\approx 10^{-3}$. From the observed probabilities, I get
$a\approx 5.4\times 10^{-12}$, as shown in Fig. \ref{fig4}, from which it follows that
$p=1.3\times 10^{-5}$.

The asymptotic formula for events in the tail of the distribution, 
Eq. (\ref{eq5}), is valid no matter how precise are $l$ and $X_{unrep}$.

{\bf Mutations and natural selection.}
Let me stress that in Fig. \ref{fig4} probabilities are measured in a set of 12 cultures. 
Thus, one expects errors of the order of $1/\sqrt{12}\approx 0.3$. In addition, 
Lenski and his group report not the occurrence of the mutation, but the moment 
at which the phenotype becomes dominant in a population. In this process, natural 
selection plays a major role.

In both the DNA-repaired and DNA-unrepaired regions of the mutation space, there exist 
points with evolutive advantage. These points act as attractors in the mutation
space.

Natural selection may be included in my model by introducing a relative fitness parameter,
$w$. \cite{ref12} $w_r=1$ and $w_u$ apply to regions of radius three around the centers of the DNA-repaired 
and DNA-unrepaired areas. Out of these regions, $w_o=0.7$. I introduce a clonal expansion phase in 
which the number of cells increases 100 times, as in the Lenski experiment, but only $N_{cell}$ 
bacteria pass to the next step. The bacteria are selected according to the conditional
probability $w/(w_o+w_r+w_u)$. Results are to be published elsewhere \cite{ref13}.

{\bf Levy model of cancer.}
With appropriate parameters, my Levy model can also be applied to mutations in stem cells and,
in particular, to the analysis of lifetime cancer risk in different tissues \cite{ref14}
with the help of a formula like Eq. (\ref{eq5}). Results are to be published elsewhere. \cite{ref15}

I would like to stress only the intriguing fact that in cases, like the ovarian germinal cell
cancer, where physical barriers act as protection, and the action of the immune system is
partially depressed, the slope $a$ takes values similar to the number obtained for bacteria.

{\bf Acknowledgments.}
The author acknowledges support from the National Program 
of Basic Sciences in Cuba, and from the Office of External Activities of the International
Centre for Theoretical Physics (ICTP).

\end{document}